\documentclass{ws-p8-50x6-00}

\def\Journal#1#2#3#4{{#1} {\bf #2}, #3 (#4)}

\def\NPB{{\em Nucl. Phys.} B}

\def\PRL{\em Phys. Rev. Lett.}
\def\PRD{{\em Phys. Rev.} D}

\def\PRC{{\em Phys. Rev.} C}
\def\APJ{{\em Astrophys. J.}}
\def\APJS{{\em Astrophys. J. Suppl.}}
\def\SCI{{\em Science}}

\def\NPA{{\em Nucl. Phys.} A}
\def\PPNP{{\em Prog. Part. Nucl. Phys.}}

\def\be{\begin{equation}}
\def\ee{\end{equation}}
\def\bea{\begin{eqnarray}}
\def\eea{\end{eqnarray}}

\def\gtwid{\mathrel{\raise.3ex\hbox{$>$\kern-.75em\lower1ex\hbox{$\sim$}}}}
\def\ltwid{\mathrel{\raise.3ex\hbox{$<$\kern-.75em\lower1ex\hbox{$\sim$}}}}

\begin{document}

\title{A STERILE NEUTRINO NEEDED FOR HEAVY-ELEMENT NUCLEOSYNTHESIS}

\author{DAVID O. CALDWELL}

\address{Physics Department, University of California,\\ Santa Barbara, CA
 93106-9530, USA\\E-mail: caldwell@slac.stanford.edu} 


\maketitle\abstracts{ A neutrino mass-mixing scheme which successfully
avoids the ``alpha effect,'' allowing $r$-process nucleosynthesis
in the neutrino-heated ejecta of supernovae, quite independently
requires the same parameters as the scheme which best fits all current
indications for neutrino mass.  The significance for particle physics is
this independent evidence for (1) at least one light sterile neutrino,
$\nu_s$; (2) a near maximally-mixed $\nu_\mu$--$\nu_\tau$ doublet split
from a lower mass $\nu_\mu$--$\nu_s$ doublet; (3) $\nu_\mu$--$\nu_e$
mixing $\gtwid 10^{-4}$; and (4) a splitting between the doublets
(measured by the $\nu_\mu$--$\nu_e$ mass difference) $\gtwid 1$ eV$^2$,
favoring the upper part of the LSND range.  If correct, it is
tantalizing that neutrinos with tiny masses which mix
with sterile species have profound effects on massive objects and the
creation of the heaviest elements.}

\section{Introduction}

While there is strong evidence\cite{ref:1} that the heaviest elements
are produced in the neutrino-heated material ejected relatively long
($\sim10$ s) after the explosion of a Type II or Type I b/c supernova,
present calculations\cite{ref:2,ref:3} show conditions which would prevent
this rapid-neutron-capture (or $r$) process from occurring.  Though
general relativistic effects\cite{ref:4} and multi-dimensional hydrodynamic
outflow\cite{ref:5} have been invoked to solve these problems, these
solutions are at best exceedingly finely tuned.  In contrast, the
solution\cite{ref:6} presented here is extremely robust, and the neutrino
mass-mixing scheme it requires is exactly that needed
if one is to explain all present evidence for neutrino mass.

In the next section, the particle physics motivations for that neutrino
scheme are discussed, with special emphasis on recent results of the LSND
experiment, since those are particularly important for the needs of the
$r$ process.  That is followed by a section on the $r$ process and its
difficulties, after which the solution is presented.

\section{Particle Physics Evidence for a Four-Neutrino Scheme}

The need for at least one light sterile (i.e., not having the usual weak
interaction) neutrino in addition to the known three active neutrinos was
proposed\cite{ref:7} as a way to explain the solar $\nu_e$ deficit, the
anomalous $\nu_\mu/\nu_e$ ratio from atmospheric neutrinos, and the apparent
need for appreciable hot dark matter.  The atmospheric anomaly, due to
$\nu_\mu\to\nu_\tau$, requires a mass-squared difference between the
$\nu_\mu$ and $\nu_\tau$ of $\Delta m^2_{\mu\tau}=0.003$ eV$^2$ and
maximal mixing ($\sin^22\theta_{\mu\tau}=1.0$), with the latter property
being quite important to solving the $r$-process problem, as shown
below.  The solar $\nu_e$ deficit is explained by $\nu_e\to\nu_s$ with
$\Delta m^2_{es}\approx10^{-5}$ eV$^2$ and a small mixing angle
($\sin^22\theta_{es}\approx0.01$) or $\Delta m^2_{es}\sim10^{-10}$
eV$^2$ and $\sin^22\theta_{es}\sim1$.  The $\nu_e$--$\nu_s$ pair is of
lower mass than the $\nu_\mu$--$\nu_\tau$ pair, which provide the hot dark
matter.  For the originally favored\cite{ref:9} critical-mass-density
universe the $\nu_\mu$ and $\nu_\tau$ needed masses of around 2 eV
each.

This phenomenology was subsequently given theoretical bases in two 1993
papers,\cite{ref:10} and has been utilized since in a large number of
publications.  Since that time the scheme has received support from the
LSND experiment, the results of which provide some measure of the mass
difference between the $\nu_e$--$\nu_s$ and $\nu_\mu$--$\nu_\tau$ pairs
and hence of the neutrino contribution to dark matter.

In its 1996 publication,\cite{ref:11} LSND claimed a signal in
$\bar\nu_\mu\to\bar\nu_e$ on the basis that 22 events of the type
$\bar\nu_ep\to e^+n$ were seen, using a stringent criterion to reduce
accidental coincidences between $e^-$ or $e^+$ and $\gamma$ rays mimicking
the 2.2-MeV $\gamma$ from $np\to d\gamma$, whereas only $4.6\pm0.06$ events
were expected.  The probability of this being a fluctuation is
$4\times10^{-8}$.  Note especially that these data were restricted to the
energy range 36 to 60 MeV to stay below the $\bar\nu_\mu$ endpoint and to
stay above the region where backgrounds are high due to the
$\nu_e\/^{12}{\rm C}\to e^-X$ reaction.  In plotting $\Delta m^2$
vs.\ $\sin^22\theta$, however, events down to 20 MeV were used to increase
the range of $E/L$, the ratio of the neutrino's energy
to its distance from the target to detection.  This plot
was intended to show the favored regions of $\Delta m^2$, and all
information about each event was used.  The likelihood analysis applied did
not have a Gaussian likelihood distribution, since its integral is infinite,
but the likelihood contour labeled ``90\%'' was obtained by going
down a factor of 10 from the maximum, as in the Gaussian case.  The
contours in the LSND plot have been widely misinterpreted as confidence
levels---which they certainly are not---because they were plotted along with
confidence-level limits from other experiments.

Recently the difficult, computer-intensive analysis in terms of real
confidence levels has been done.\cite{ref:12}  The likelihood for a grid in
($\sin^22\theta$, $\Delta m^2$) space, including backgrounds, has been computed
and compared with numerous Monte Carlo experiments to obtain a 90\% confidence
region.  While the equivalency varies from point to point in the
$\Delta m^2$--$\sin^22\theta$ plane, a typical value for the 90\%
confidence level is
 down a factor of 20 from the likelihood
maximum.  Thus the LSND allowed regions are considerably broader in
$\sin^22\theta$ than in the plots published so far, and other experiments
constrain allowed $\Delta m^2$ regions less.

The confusion of comparing likelihood levels for LSND with confidence levels
from other experiments may be exacerbated by using the 20--36 MeV region
for the LSND data.  While this higher background energy range makes some
difference for the 1993--5 data, it could have had an appreciable effect for
the parasitic 1996--8 runs, which were at a low event rate, increasing the
effect of cosmic ray background.  This could raise the low end of the
supposed signal energy spectrum, especially as the one LSND distribution
which was statistically worrisome was the ratio ($R$) of real to accidental
events.  Some accidental events in this 20--36 MeV region would favor
low values of $\Delta m^2_{e\mu}$ making the higher $\Delta m^2$ values
desirable for dark matter appear less likely.

\begin{figure}[t]
\epsfxsize=7cm 
\epsfbox{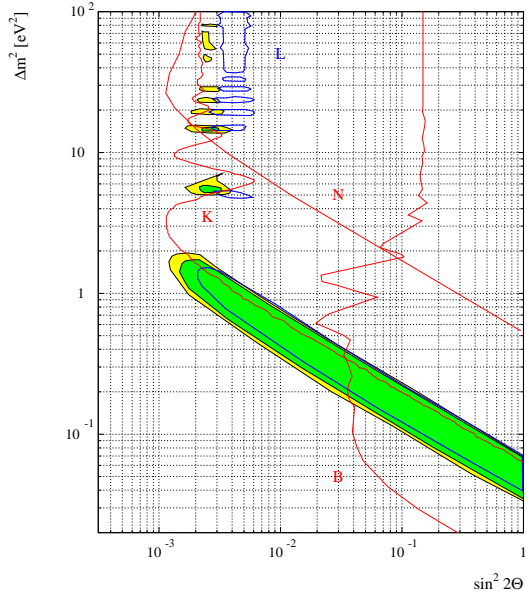} 
\caption{Filled in areas are 90\% and 95\% confidence regions based on
the product of the KARMEN and LSND Feldman-Cousins likelihood ratios.  Also
shown is the Feldman-Cousins 90\% confidence region for LSND alone (``L'').
Left of the ``K'', ``N'', and ``B'' curves are exclusion regions of
KARMEN, NOMAD, and Bugey.\label{fig:1}}
\end{figure}

Nevertheless, when a joint analysis\cite{ref:12} is made of the LSND and
KARMEN\cite{ref:13}
experiments even using the 20--36 MeV range for LSND, the region around 5.5
eV$^2$ is as probable as the banana-shaped region at lower $\Delta m^2$, as
shown in Fig.~\ref{fig:1}.
Frequently ignored by theorists, this higher mass region is favored by the
$\nu_\mu\to\nu_e$ LSND data.  Of course in the
$\nu_\mu\to\nu_e$ case,\cite{ref:11} using $\nu_\mu$ from $\pi^+$ decay in
flight and detecting $\nu_e$ by $\nu_e\/^{12}{\rm C}\to e^-X$, the
backgrounds are higher and hence yield much poorer statistics than for
$\bar\nu_\mu\to\bar\nu_e$ with $\bar\nu_\mu$ from $\mu^+$ at rest.  In
addition to the $\Delta m^2$ issue, the important point of Fig.~\ref{fig:1}
is that although the KARMEN data are consistent with background, the joint
analysis of the $\bar\nu_\mu\to\bar\nu_e$ data from the two experiments shows
an appreciable region for a signal.  KARMEN is continuing to take data, and
LSND will have an improved analysis available soon.  This new analysis
has produced an
excellent $R$ distribution for all the data and an energy distribution
with reduced contributions at the low end, favoring higher $\Delta m^2_{e\mu}$
values than does Fig.~\ref{fig:1}.

\section{Problems with Synthesis of the Heaviest Elements}

While in the next section we will find that the $r$ process of rapid
neutron capture in supernovae provides strong support for the double doublet
of neutrinos, initially the reverse appeared to be true, with the $r$
process apparently placing stringent limits on $\nu_\mu$--$\nu_e$ mixing.
The origin of these limits is that energetic $\nu_\mu$
($\langle E\rangle\approx25$ MeV) coming from deep in the supernova core
could convert via an MSW transition to $\nu_e$ inside the region of the
$r$-process, producing $\nu_e$ of much
higher energy than the thermal $\nu_e\ (\langle E\rangle\approx11$ MeV).  The
latter, because of their charged-current interactions, emerge from farther out
in the supernova where it is cooler.  Since the cross section for $\nu_en\to
e^-p$ rises as the square of the energy, these converted energetic $\nu_e$
would deplete neutrons, stopping the $r$-process. 
Calculations\cite{ref:14} of this effect limit $\sin^22\theta$ for
$\nu_\mu\to\nu_e$ to $\ltwid10^{-4}$ for $\Delta m^2_{e\mu}\gtwid2$ eV$^2$, in
conflict with at least the higher mass region of the LSND results, which will
be of particular interest here.

More recently, serious problems have been found with the $r$ process itself.
First, simulations\cite{ref:2} have revealed the $r$-process
region to be insufficiently neutron-rich, since about $10^2$ neutrons is
required for each seed nucleus, such as iron.  This was bad enough, but the
recent realization of the full effect of $\alpha$-particle formation has
created a disaster for the $r$ process.\cite{ref:3}  At a radial region
inside where the $r$ process should occur, all available protons swallow
up neutrons to form the very stable $\alpha$ particles, following which
$\nu_en\to e^-p$ reactions reduce the neutrons further and create more protons
which make more $\alpha$ particles, and so on.  The depletion of neutrons
rapidly shuts off the
$r$ process, and essentially no nuclei above $A=95$ are produced.

To solve this problem the $\nu_e$ flux has to be removed before the
$r$ process site, while leaving a very large $\nu_e$ flux at a
smaller radius for material heating and ejection.  The obvious difficulty
of accomplishing this has led to searches for other possible sites for the
$r$ process, such as neutron star mergers.

\section{Neutrino Solution for a Successful $r$ Process}

The apparent miracle of having a huge $\nu_e$ flux disappear before it
reaches the radius of the supernova where $\alpha$ particles form can be
accomplished\cite{ref:6} if there is (1) a sterile neutrino, (2)
approximately maximal $\nu_\mu\to\nu_\tau$ mixing, (3) $\nu_\mu\to\nu_e$
mixing $\gtwid10^{-4}$, and (4) an appreciable ($\gtwid1$ eV$^2$) mass-squared
difference
between $\nu_s$ and the $\nu_\mu$--$\nu_\tau$.  This is precisely the
neutrino mass pattern required to explain the solar and atmospheric anomalies
and the LSND result, plus providing some hot dark matter!

Such a mass-mixing pattern creates two level crossings.  The inner one,
which is outside the neutrinosphere (beyond which neutrinos can readily
escape) is near where the $\nu_{\mu,\tau}$ potential $\propto(n_{\nu_e}-n_n/2)$
goes to zero.  Here $n_{\nu_e}$ and $n_n$ are the numbers of $\nu_e$ and
neutrons, respectively.  The $\nu_{\mu,\tau}\to\nu_s$ transition which occurs
depletes the dangerous high-energy $\nu_{\mu,\tau}$ population.  Outside
of this level crossing, another occurs where the density is appropriate for a
matter-enhanced MSW transition corresponding to whatever $\Delta m^2_{e\mu}$
LSND is observing.  Because of the $\nu_{\mu,\tau}$ reduction at the first
level crossing, the dominant process in the MSW region reverses from the
deleterious $\nu_{\mu,\tau}\to\nu_e$, becoming $\nu_e\to\nu_{\mu,\tau}$ and
dropping the $\nu_e$ flux.  For an
appropriate value of $\Delta m^2_{e\mu}$, the two level crossings are
separate but sufficiently close so that the transitions are coherent.  Then
with adiabatic transitions (as calculations show) and maximal
$\nu_\mu$--$\nu_\tau$ mixing, the neutrino flux emerging from the second
level crossing is 1/4 $\nu_\mu$, 1/4 $\nu_\tau$, and 1/2 $\nu_s$, and no
$\nu_e$.

A more exact way to explain this in the four-neutrino formalism is to
transform the four mass eigenstates into the flavor states via the
mixing angles $\phi$ for solar $\nu_e\to\nu_s$, $\omega$ for LSND
$\nu_\mu\to\nu_e$, and $\pi/4$ (maximal mixing) for atmospheric
$\nu_\mu\to\nu_\tau$.  Symbolically,
$$\pmatrix{|\nu_s\rangle \cr
           |\nu_e\rangle \cr
           |\nu_\mu\rangle \cr
           |\nu_\tau\rangle \cr}
= \pmatrix{\hbox{Atmospheric} \cr
           \nu_\mu\to\nu_\tau \cr}
  \pmatrix{\hbox{Solar} \cr
           \nu_e\to\nu_s \cr}
  \pmatrix{\hbox{LSND} \cr
           \nu_\mu\to\nu_e \cr}
  \pmatrix{|\nu_1\rangle \cr
           |\nu_2\rangle \cr
           |\nu_3\rangle \cr
           |\nu_4\rangle \cr},
$$
giving
$$\pmatrix{|\nu_e\rangle \cr
           |\nu_s\rangle \cr
           |\nu^*_\mu\rangle \cr
           |\nu^*_\tau\rangle \cr}
= \pmatrix{ \cos\phi & \sin\phi\cos\omega & \sin\phi\sin\omega & 0 \cr
            -\sin\phi & \cos\phi\cos\omega & \cos\phi\sin\omega & 0 \cr
            0 & -\sin\omega & \cos\omega & 0 \cr
            0 & 0 & 0 & 1 \cr}
  \pmatrix{|\nu_1\rangle \cr
           |\nu_2\rangle \cr
           |\nu_3\rangle \cr
           |\nu_4\rangle \cr},
$$
where
\begin{eqnarray*}
|\nu^*_\mu\rangle&=1/\sqrt2(|\nu_\mu\rangle-|\nu_\tau\rangle)
=-\sin\omega|\nu_2\rangle+\cos\omega|\nu_3\rangle\\
|\nu^*_\tau\rangle&=1/\sqrt2(|\nu_\mu\rangle+|\nu_\tau\rangle)
=|\nu_4\rangle,\ \hbox{a mass eigenstate}.
\end{eqnarray*}

In this formalism what occurs at the first level crossing is
$\nu^*_\mu\to\nu_s$,
and at the second, $\nu_e\to\nu^*_\mu$, while $\nu^*_\tau$ being a mass
eigenstate goes through both regions unaffected.  Again this gives
1/4 $\nu_\mu$, 1/4 $\nu_\tau$, 1/2 $\nu_s$, and no $\nu_e$ at all.

Note that the $\bar\nu_e$ flux is also unaffected at the level crossings,
so $\bar\nu_ep\to e^+n$ enhances the neutron number in the $r$ process region,
since the protons have not been depleted by $\alpha$ particle formation.
It should be emphasized that this mechanism is quite robust,
not depending on details of the supernova dynamics, especially as it occurs
quite late in the explosive expansion.

It is essential that the two level crossings be in the correct order, and this
provides a requirement on $\Delta m^2_{e\mu}$, since the MSW transition
depends on density and hence on radial distance from the protoneutron star.
Detailed calculations have been made for $\Delta m^2_{e\mu}\sim6$ eV$^2$,
which works very well.  Possibly $\Delta m^2_{e\mu}$ as low as 2 eV$^2$ or
maybe even 1 eV$^2$ would work, but that is speculative.  At any rate, the
mass difference needed in this scheme, which is the only one surely consistent
with all manifestations of neutrino mass and which rescues the $r$
process,\cite{ref:15} implies appreciable hot dark matter.

\section{Conclusions}

It is quite remarkable that the profound problems of producing the
heaviest elements by supernovae can be solved in a manner which requires
no adjustment of parameters if the arrangement of masses and mixings of
neutrinos is exactly that required to explain the solar $\nu_e$ deficit,
the atmospheric neutrino anomaly, and the observations of the LSND experiment
(or alternatively the need for hot dark matter).  This is achieved via an
active-sterile level crossing in the supernova, followed by an active-active
transition.  The total independence of this supernova information strongly
enhances the case for this four-neutrino scheme.

\section*{Acknowledgments}

This paper is based largely on work done with G.M.~Fuller and Y.-Z.~Qian,
to whom I am grateful, and was supported in part by the U.S.~Department of
Energy under contract DE-FG03-91ER40618.


\end{document}